\documentstyle[eqsecnum,preprint,aps,tighten,psfig]{revtex}

\begin{document}
\draft

\title{Life time of resonances in transport simulations}

\author{Stefan Leupold}

\address{Institut f\"ur Theoretische Physik, Justus-Liebig-Universit\"at
Giessen,\\
D-35392 Giessen, Germany}


\maketitle

\begin{abstract}
We calculate the life time of a resonance in our recently developed framework
for a test-particle description of transport processes for states with 
continuous mass spectra. The result differs from the expression commonly used
in transport simulations but agrees with the one derived by Danielewicz and Pratt
relating the life time to the scattering phase shift.
\end{abstract}

\section{Introduction}

For the understanding of heavy-ion collisions semi-classical transport theory 
has become an indispensable tool (cf.~e.g.~\cite{BeDG88,CMMN90}). The theoretical
foundation of this approach goes back to the pioneering works on
non-equilibrium quantum field theory \cite{Sc61,KB,BM63,Ke64}. While former works 
have focused their attention more or less on the quasi-particle regime 
(see e.g.~\cite{Da84a,Ch85,BotMal,DanMrow,MrHe,Mrow97,GrLe98} and references 
therein) the extension of the formalism to off-shell phenomena has become a topic 
of growing interest in the last few years 
\cite{henning,bozek97,bozek99,IKV99,EBM99,EM99,CJ99,leupold00,CJ9923} 
since it has been realized that the
collision rates present in high energetic nucleus-nucleus collisions typically are
so large that an on-shell approximation seems to be inappropriate. In addition, the 
resonances excited during the
reaction may have large decay widths. Therefore, a representation of these states by
stable particles may not be a proper approximation. 

The usual approach to solve a transport equation is the representation of the
phase-space density by test-particles \cite{BeDG88,CMMN90} (concerning off-shell
extensions see \cite{EBM99,EM99,CJ99,leupold00,CJ9923}). If not only asymptotically
stable states but also resonances are simulated by test-particles, one has to 
attribute
a finite life time to those `resonance test-particles', i.e.~one has to demand that
they decay after some time of propagation --- provided that they had not suffered
from a collision prior to their decay. Such resonance test-particles usually
have arbitrary invariant masses (chosen according to their spectral distribution).
The question which (in general energy- or mass-dependent) life time has to be 
attributed to the resonance test-particles is under present discussion 
\cite{DanPratt}. The commonly used recipe is to identify the life time with the
inverse decay width of the resonance, evaluated for the respective invariant mass.
Near threshold the width thus becomes small due to the available phase-space. Hence
the life time --- if identified with the inverse width --- becomes large. In
\cite{DanPratt} it was suggested to rather calculate the life time from the
time delay that the particles suffer which form the resonance. This time delay
is given by the energy derivative of the phase shift measured for the particles 
which scatter by forming the resonance as an intermediate state. This quantity
--- and therefore also the life time calculated in this way --- vanishes near
threshold. Hence the two expressions for the life time show a completely different
behavior as functions of the invariant mass of the resonance.

Recently a novel approach to the test-particle description of transport processes 
for states with a continuous mass spectrum has been presented in 
\cite{leupold00,leupKBproc}. There the formalism was outlined for a system of
non-relativistic (asymptotically stable) states subject to elastic collisions. 
In principle, however, there is no obstacle to treat also resonances in the same
framework and to answer the question about the proper life time for such states. 
Since in principle the creation and decay of particles is a matter of relativistic 
field theory we find it more appropriate to present the formalism here for a
relativistic system. One of the goals of \cite{leupold00} was the
derivation of the (non-relativistic) equations of motion for the test-particles. We
do not repeat the straightforward generalization for the relativistic case here. 
The resulting relativistic
equations of motion are given in \cite{CJ99} for energy-momentum independent
self-energies and in \cite{CJ9923} for arbitrary self-energies. We note that the
way in which these equations are obtained in \cite{CJ99,CJ9923} is different
from our approach \cite{leupold00}. Nonetheless the resulting equations of
motion are the same. (This issue is discussed in detail in \cite{leupold00}.)

In the next Section we generalize the approach of \cite{leupold00} to
a relativistic system (of scalar bosons). In this way we set the stage for a proper
treatment of resonances and automatically also review the most
important points of the formalism outlined in \cite{leupold00}. 
In Sec.~\ref{sec:lifetime} we calculate
the life time of resonances for various situations. We summarize our results in 
Sec.~\ref{sec:sum}. An Appendix is added which contains some technical details.

\section{Reviewing the derivation of the transport equation and the effective
particle number density}

\begin{figure}[htbp]
\centerline{\psfig{figure=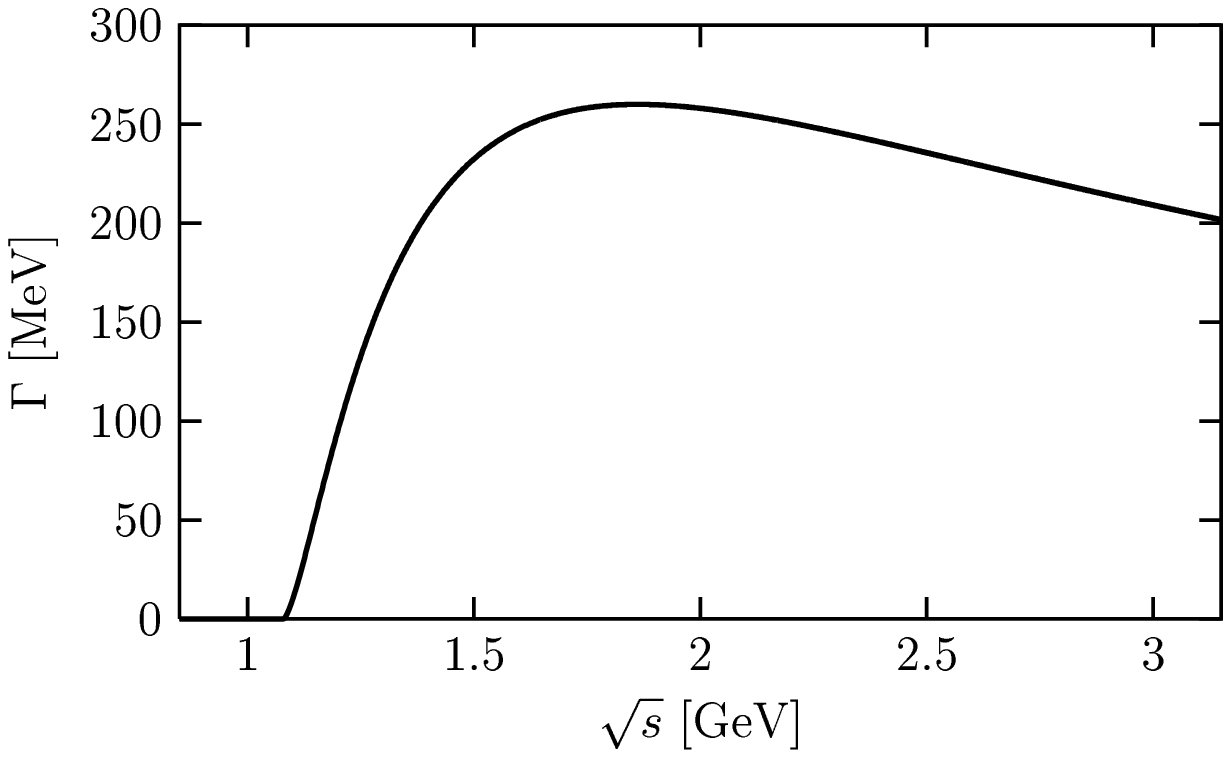}}
\caption{Width $\Gamma$ as given in (\protect\ref{eq:gamdelta})
as a function of the invariant mass.}
\label{fig:width}
\end{figure}

\begin{figure}[htbp]
\centerline{\psfig{figure=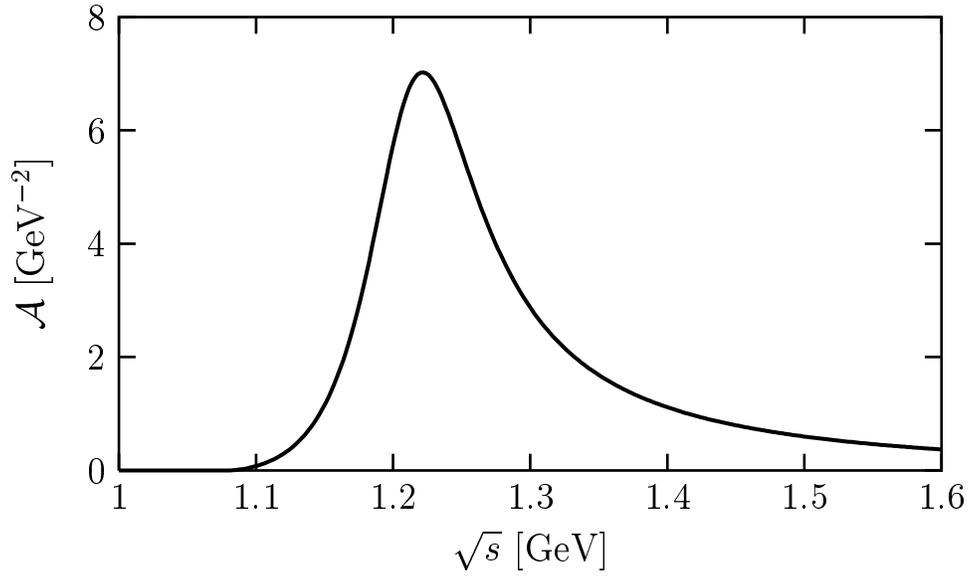}}
\caption{Spectral function ${\cal A}$ as given in (\protect\ref{eq:defspec})
as a function of the invariant mass.}
\label{fig:spectral}
\end{figure}

We start out from the Kadanoff-Baym equations \cite{KB} for relativistic scalar 
fields (cf.~e.g.~\cite{GrLe98,CJ99} and references therein)
\begin{eqnarray}
  \label{eq:exeom1}
(-\Box_1 -m^2) D^<(1,1') &=& 
\int \!\! d\bar 1 
\left[
    \Sigma^{\rm ret}(1,\bar 1) \, D^<(\bar 1, 1') 
  + \Sigma^<(1,\bar 1) \, D^{\rm av}(\bar 1, 1')
\right]  
\,, \\
  \label{eq:exeom2}
(-\Box_1 -m^2) D^>(1,1') &=& 
\int \!\! d\bar 1 
\left[ 
    \Sigma^{\rm ret}(1,\bar 1) \, D^>(\bar 1, 1') 
  + \Sigma^>(1,\bar 1) \, D^{\rm av}(\bar 1, 1')
\right]  
\end{eqnarray}
where we have introduced the two-point functions without ordering
\begin{eqnarray}
  \label{eq:defdless}
i D^< (x,y) &=& \langle \phi(y) \,\phi(x) \rangle  \,, \\
  \label{eq:defdmore}
i D^> (x,y) &=& \langle \phi(x) \,\phi(y) \rangle 
\end{eqnarray}
and the retarded and advanced quantities
\begin{eqnarray}
  \label{eq:defdret}
F^{\rm ret}(x,y) &=& \Theta(x_0-y_0) \left[ F^>(x,y) -F^<(x,y) \right] \,, \\
  \label{eq:defdav}
F^{\rm av}(x,y) &=& \Theta(y_0-x_0) \left[ F^<(x,y) -F^>(x,y) \right]  
\end{eqnarray}
with $F = D,\Sigma$. The self-energy is denoted by $\Sigma$.
After Wigner transformation 
\begin{equation}
\bar F(X,p) = \int \!\! d^4\!u \, e^{ipu} F(X+u/2,X-u/2)  
\end{equation}
of all quantities a gradient expansion of the Kadanoff-Baym equations is performed.
Next one neglects all contributions which are effectively of second or higher order
in the derivative with respect to the center-of-mass variable $X$. In this way one
ends up with a transport equation (cf.~\cite{CJ99} for details of the relativistic
case, the corresponding non-relativistic case is discussed in detail in 
\cite{leupold00})
\begin{equation}
  \label{eq:transport1}
\sqrt{s} \Gamma {\cal A}\, [p^2-m^2-{\rm Re}\bar \Sigma^{\rm ret},S^<]
-{\cal A}\, [\sqrt{s} \Gamma , (p^2-m^2-{\rm Re}\bar \Sigma^{\rm ret}) S^<]
= \sqrt{s} \Gamma \, S^< - i \bar\Sigma^< {\cal A} 
\end{equation}
where we have introduced the following real-valued quantities: \\
two-point functions
\begin{equation}
  \label{eq:defskl}
S^{<,>}(X,p) = i \bar D^{<,>}(X,p)  \,,
\end{equation}
spectral function\footnote{Note that the definition of the spectral function
slightly differs from the non-relativistic version \cite{leupold00}. In the 
non-relativistic limit there is the following connection between relativistic
and non-relativistic quantities: ${\cal A}_{\rm rel} \to {\cal A}_{\rm nr}/(4m)$,
$\Gamma_{\rm rel} \to \Gamma_{\rm nr}$.} ($s = p^2$)
\begin{equation}
  \label{eq:defspec}
{\cal A}(X,p) = {1 \over 2} \left( S^>(X,p) - S^<(X,p) \right) = 
{\sqrt{s} \Gamma(X,p) \over 
 (p^2-m^2-{\rm Re}\bar \Sigma^{\rm ret}(X,p))^2 + s\Gamma^2(X,p) }  \,,
\end{equation}
width
\begin{equation}
  \label{eq:defwidth}
\Gamma(X,p) = {1 \over \sqrt{s}} 
{i\over 2} \left(\bar\Sigma^>(X,p)- \bar\Sigma^<(X,p) \right)  \,.
\end{equation}
A typical example for the width and the corresponding spectral function
is depicted in Figs.~\ref{fig:width} and \ref{fig:spectral}, respectively,
for a toy model displaying the main features of the $\Delta$ baryon. Width and
spectral function start to deviate from zero at threshold (here pion plus nucleon 
mass). The spectral function grows until the nominal resonance mass is reached. 
The width grows beyond that point. The large energy behavior is chosen such that
$\sqrt{s}\Gamma$ finally reaches a constant value. (Note that the displayed mass 
range for $\Gamma$ is larger than the one for the spectral function.)
Details of the calculation for $\Gamma$ (and of the corresponding real part of the 
retarded self-energy) are deferred to the Appendix. They
form the basis of all the plots which exemplify the relevant quantities.

\begin{figure}[htbp]
\centerline{\psfig{figure=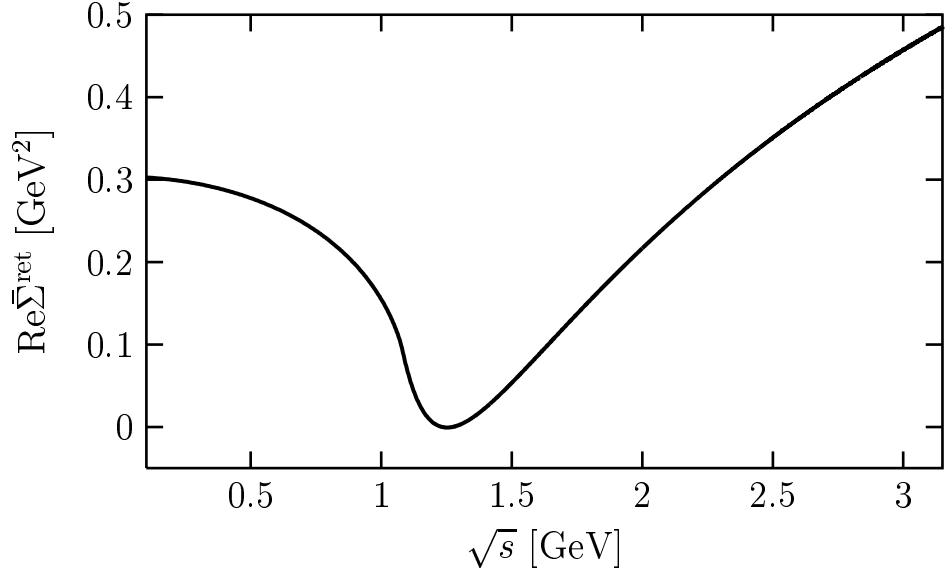}}
\caption{Real part of the self-energy as given in (\protect\ref{eq:dispexa})
as a function of the invariant mass.}
\label{fig:resigma}
\end{figure}

It is worth noting that the spectral function is normalized
\begin{equation}
  \label{eq:norm}
\int\limits_{-\infty}^{+\infty} \! {dp_0 \over \pi} \, p_0 \, {\cal A} = 
\int\limits_0^{\infty}  {d(p_0^2) \over \pi} \, {\cal A} = 1   \,,
\end{equation}
if the real part of the retarded self-energy is connected to the width (imaginary
part) via a (in general $n$-times subtracted) dispersion relation
\begin{equation}
  \label{eq:dispersion}
{\partial^n \over \partial p_0^n}
{\rm Re}\bar\Sigma^{\rm ret}(X;p_0,\vec p) = 
{\partial^n \over \partial p_0^n}
{\cal P} \int\!{dk_0 \over \pi} \, {1 \over p_0 -k_0} \, \sqrt{k_0^2 -\vec p^2} \,
\Gamma(X;k_0,\vec p)
\end{equation}
where ${\cal P}$ denotes the principal value. An example for the real part of the 
retarded self-energy is shown in Fig.~\ref{fig:resigma} (cf.~the Appendix for 
details). 

The generalized Poisson bracket used in (\ref{eq:transport1}) is defined as
\begin{equation}
[A,B] = \partial_{X_0} A \,\partial_{p_0} B - \partial_{p_0} \,A \partial_{X_0} B 
- \vec\nabla_X A \,\vec\nabla_p B + \vec\nabla_p A \,\vec\nabla_X B  \,.
\end{equation}
Note that the drift term inherent to any kinetic equation is hidden in 
\begin{equation}
  \label{eq:drift}
[p^2-m^2,\, B] = -2 p_0 \partial_{X_0} B - 2 \vec p \cdot \vec \nabla_X B   \,.
\end{equation}
The r.h.s.~of (\ref{eq:transport1}) can be expressed via the common gain and loss
terms using the definitions (\ref{eq:defspec}) and (\ref{eq:defwidth}):
\begin{equation}
  \label{eq:gainloss}
\sqrt{s} \Gamma \, S^< - i \bar\Sigma^< {\cal A} = 
\underbrace{{i \over 2} \bar\Sigma^>\,S^<}_{\rm loss} - 
\underbrace{{i \over 2} \bar\Sigma^<\,S^>}_{\rm gain}  \,.
\end{equation}

In the most general case of an interacting multi-particle state away from thermal
equilibrium, the transport equation (\ref{eq:transport1}) is a rather involved
integro-differential equation. Within the quasi-particle approximation a quite 
successful method to solve the corresponding transport equation is to use a 
test-particle ansatz for the phase-space density \cite{BeDG88,CMMN90}. 
From the drift and (generalized) Vlasov terms of the transport
equation one deduces the equations of motion for the test-particles while the 
scattering integrals yield the appropriate cross-sections for the
collisions among the test-particles. For the case at hand, i.e.~beyond the 
quasi-particle approximation,
it is suggestive to use also a test-particle representation
\begin{equation}
  \label{eq:testpart}
f(t,\vec x;p) \sim 
\sum\limits_i \delta^{(3)}(\vec x - \vec x_i(t)) \, 
\delta(p^2_0 - E^2_i(t)) \,
\delta^{(3)}(\vec p - \vec p_i(t))    \,.
\end{equation}
Note that here in contrast to the quasi-particle approximation the test-particles
are allowed to have arbitrary energies {\em not} connected to their respective
three-momentum by any on-shell condition. The remaining question is which quantity
$f$ should be represented by test-particles. 

So far we have reviewed the derivation of the transport equation 
(\ref{eq:transport1}) as outlined in more detail in \cite{CJ99}. Albeit starting 
from the same transport equation
we differ from the approach of \cite{CJ99,CJ9923} in what follows concerning the 
interpretation and use of this equation. In \cite{CJ99,CJ9923} the transport 
equation was merely used as a tool to determine the evolution of test-particles
{\em between} collisions. In the approach presented here and in \cite{leupold00}
we adopt the point of view that the transport equation (\ref{eq:transport1})
once solved as exactly as possible provides the appropriate propagation, scattering
and decays of the test-particles. 
The differences between \cite{CJ99} and our approach are discussed in more detail in
\cite{leupold00}. One difference concerns exactly the question which quantity $f$ 
should be represented by test-particles. 

There is an obvious constraint for such
a quantity: the corresponding number of test-particles
\begin{equation}
  \label{eq:numtest}
N_{tp} = \int\!\! d^3\!x \int\!\!{d^4\!p \over (2\pi)^4} \,p_0 \, f(t,\vec x;p)
\end{equation}
has to remain constant during the propagation of the test-particles and also
for elastic scattering events encoded by $\bar\Sigma^{<,>}_{\rm elast}$. 
Only inelastic scatterings and particle decays both contained in 
\begin{equation}
  \label{eq:inelst}
\bar\Sigma^{<,>}_{\rm inelast} = \bar\Sigma^{<,>} - \bar\Sigma^{<,>}_{\rm elast}
\end{equation}
should change the number of test-particles $N_{tp}$. For the following 
considerations we need an important property of the elastic scattering part of the
self-energies which is a consequence of detailed balance:
\begin{equation}
  \label{eq:elastdetbal}
\int \!\!{d^4\!p \over (2\pi)^4} 
\left( i \bar\Sigma^<_{\rm elast} S^> - i \bar\Sigma^>_{\rm elast} S^< \right)
= 0  \,.
\end{equation}
A natural choice for $f$ seems to be
\begin{equation}
  \label{eq:wrongchoice}
f(t,\vec x;p) \stackrel{?}{=} S^<(t,\vec x;p) \,.
\end{equation}
Indeed, the full Kadanoff-Baym equations (\ref{eq:exeom1}, \ref{eq:exeom2})
conserve (\ref{eq:numtest}) with the choice (\ref{eq:wrongchoice}) as long as
only elastic interactions are present. (This was the reason for the choice 
(\ref{eq:wrongchoice}) in \cite{CJ99,CJ9923}.)
However, we no longer deal with the full
Kadanoff-Baym equations but with the transport equation (\ref{eq:transport1})
obtained from the former by gradient expansion. 
From (\ref{eq:transport1}) it is straightforward to get
\begin{eqnarray}
&&
{d \over dt} \int\!\! d^3\!x \int\!\!{d^4\!p \over (2\pi)^4} \,p_0 \,S^<(t,\vec x;p)
\nonumber \\ && 
= {d \over dt} \int\!\! d^3\!x \int\!\!{d^4\!p \over (2\pi)^4} \, p_0 \, K \, S^< 
+ \int\!\! d^3\!x \int\!\!{d^4\!p \over (2\pi)^4} \, 
{1 \over 4 \sqrt{s}\Gamma \, {\cal A} } 
\left( i \bar\Sigma^< S^> - i \bar\Sigma^> S^< \right)  
  \label{eq:timederwc}
\end{eqnarray}
with
\begin{equation}
  \label{eq:defk}
K = {1 \over 2 p_0} 
\left( 
{\partial {\rm Re}\bar\Sigma^{\rm ret} \over \partial p_0} + 
{p^2 - m^2 - {\rm Re}\bar\Sigma^{\rm ret} \over \sqrt{s}\Gamma}
{\partial(\sqrt{s}\Gamma) \over \partial p_0}
\right)   \,.
\end{equation}
Even if the self-energies are restricted to their elastic parts, the r.h.s.~of 
(\ref{eq:timederwc}) {\em does not vanish}. In contrast, for the choice
\begin{equation}
  \label{eq:defskleff}
f = S^<_{\rm eff} := 2 \sqrt{s}\Gamma \, {\cal A}\,(1-K)\,S^<
\end{equation}
we deduce from (\ref{eq:transport1}) (cf.~\cite{leupold00})
\begin{equation}
  \label{eq:timederrc}
{d \over dt} \int\!\! d^3\!x \int\!\!{d^4\!p \over (2\pi)^4} \,p_0 
\,S^<_{\rm eff}(t,\vec x;p)
= \int\!\! d^3\!x \int\!\!{d^4\!p \over (2\pi)^4} \,{1 \over 2} \,
\left( i \bar\Sigma^< S^> - i \bar\Sigma^> S^< \right) \,.
\end{equation}
Using (\ref{eq:elastdetbal}) we find that this time derivative only deviates from 
zero in the presence of inelasticities. Hence we conclude that (\ref{eq:defskleff})
is a proper choice for the test-particle number density. Note that the transport
equation (\ref{eq:transport1}) constitutes an effective theory derived
from the underlying full quantum field theory by gradient expansion. In this
approximation scheme the parts of $S^<(X,p)$ with fast oscillations in $X$ are
neglected. It is not hard to imagine that such a neglect necessitates the 
redefinition of conserved quantities as --- in our case at hand --- the particle 
number density. In the following we will refer to $S^<_{\rm eff}$ as the effective
particle number density. 

It is fortunate to rewrite the transport equation in terms of this quantity:
\begin{eqnarray}
&& [p^2-m^2-{\rm Re}\bar \Sigma^{\rm ret},{S_{\rm eff}^< \over 1-K}]
-{1 \over \sqrt{s} \Gamma}\, 
[\sqrt{s} \Gamma , (p^2-m^2-{\rm Re}\bar \Sigma^{\rm ret}) \,
{S_{\rm eff}^< \over 1-K}]  \nonumber \\ &&
= {i\bar\Sigma^> \over 2 \sqrt{s}\Gamma \, {\cal A}\,(1-K)} \,S_{\rm eff}^< 
- i\bar\Sigma^<\,S^>  
  \label{eq:transport2}
\end{eqnarray}
with
\begin{equation}
  \label{eq:sgrskleff}
S^> = 2{\cal A} \,b 
\end{equation}
and the Bose enhancement factor
\begin{equation}
  \label{eq:bose}
b = 1 + {S_{\rm eff}^< \over 4 \sqrt{s}\Gamma \, {\cal A}^2\,(1-K)}  \,.
\end{equation}
To obtain (\ref{eq:sgrskleff}) we have expressed $S^>$ in terms of ${\cal A}$
and $S^<$ according to (\ref{eq:defspec}). $S^<$ is expressed in terms of
$S_{\rm eff}^<$ using (\ref{eq:defskleff}).
Note that in (global) thermal equilibrium the following relations hold:
\begin{equation}
  \label{eq:thermal}
\left.  
\begin{array}{rcl}
S^<(p) &=& 2 n_B(p_0)\,{\cal A}(p)  \\ & & \\
S_{\rm eff}^<(p) &=& 2 n_B(p_0)\, 2 \sqrt{s}\Gamma(p) \, {\cal A}^2(p)\,(1-K(p)) \\
& & \\
i\bar\Sigma^<(p) &=& 2 n_B(p_0)\,\sqrt{s}\Gamma(p)  \\ & & \\
b(p) &=& 1 + n_B(p_0)
\end{array}
\right\} \mbox{thermal equilibrium}
\end{equation}
where we have introduced the Bose distribution $n_B$.

The transport equation (\ref{eq:transport2}) forms the basis of the following
considerations.

\section{Life time of resonances}   \label{sec:lifetime}

\subsection{Vacuum case}

\begin{figure}[htbp]
\centerline{\psfig{figure=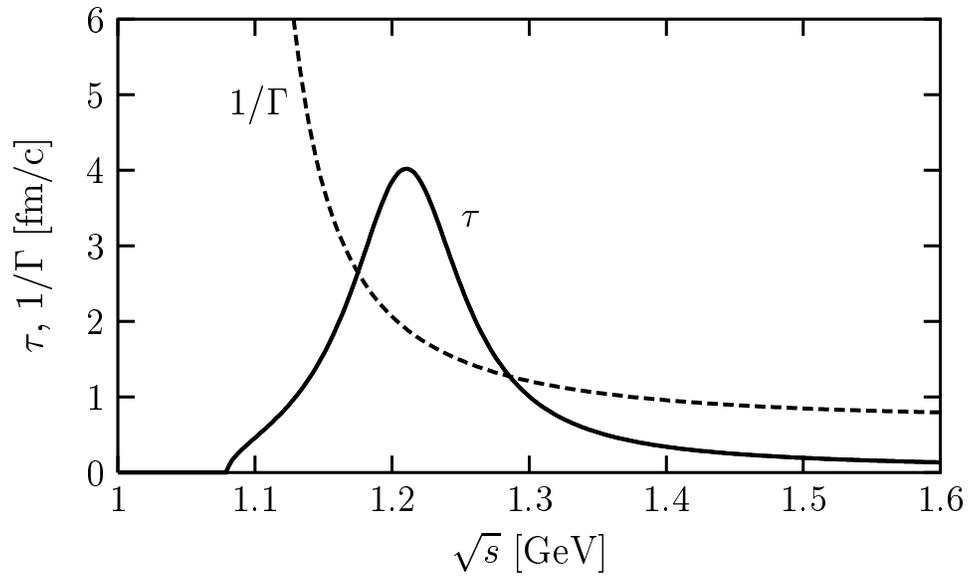}}
\caption{Life time $\tau$ (full line) as given in (\protect\ref{eq:lifetime})
and $1/\Gamma$ (dashed line)
as functions of the invariant mass. Note that $\tau$ is evaluated in the rest 
frame of the decaying resonance, i.e.~$p_0 = \sqrt{s}$.}
\label{fig:tau}
\end{figure}

\begin{figure}[htbp]
\centerline{\psfig{figure=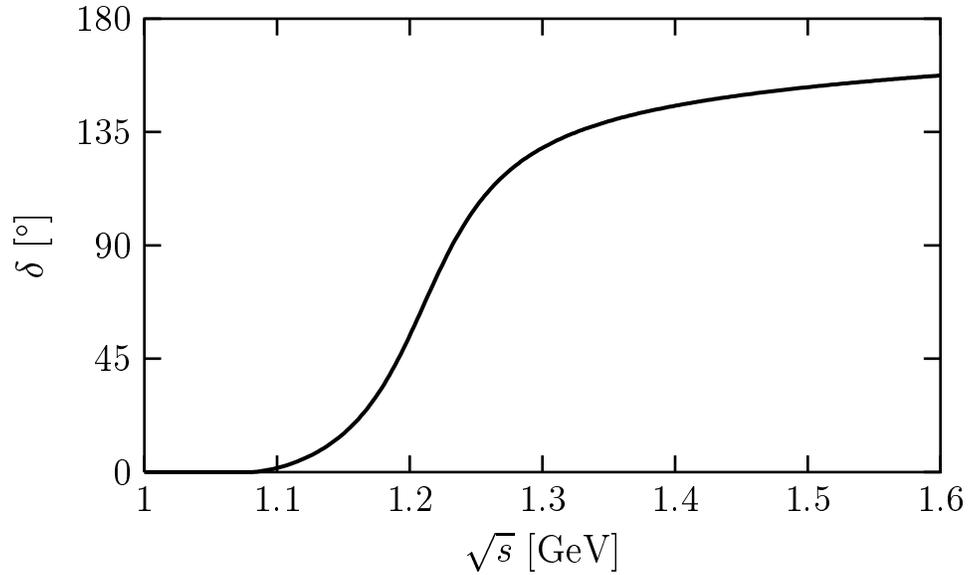}}
\caption{Phase shift $\delta$ as defined in (\protect\ref{eq:defphaseshift})
as a function of the invariant mass.}
\label{fig:phase}
\end{figure}

For the most general off-equilibrium situation 
the transport equation (\ref{eq:transport2}) is rather complicated.
To focus on the aspect relevant for our purpose we restrict ourselves to the 
following scenario: A bunch of resonances (which do not interact 
with each other) with invariant mass 
$\sqrt{s} = \sqrt{p^2}$ is uniformly distributed in vacuum and their
decay time is determined. This allows for drastic simplifications of the transport
equation. First of all, any dependence on $\vec x$ vanishes. Second, the gain rate
$i\bar\Sigma^<$ vanishes since no new resonances are formed. Finally, the 
self-energies do not depend on time. Thus we get 
\begin{equation}
  \label{eq:intermed}
-\left(2 p_0 - {\partial {\rm Re}\bar\Sigma^{\rm ret} \over \partial p_0} \right)
{1 \over 1-K} \, \partial_t S^<_{\rm eff} + 
{\partial(\sqrt{s}\Gamma) \over \partial p_0}
{p^2 - m^2 - {\rm Re}\bar\Sigma^{\rm ret} \over \sqrt{s}\Gamma \, (1-K)} 
\partial_t S^<_{\rm eff} 
= {i\bar\Sigma^> \over 2 \sqrt{s}\Gamma \, {\cal A}\,(1-K)} \,S_{\rm eff}^< \,.
\end{equation}
Using (\ref{eq:defwidth}) (note: $i\bar\Sigma^< = 0$)
and (\ref{eq:defk}) one ends up with the very simple relation
\begin{equation}
  \label{eq:timeevsimple}
- 2 p_0 \, \partial_t S^<_{\rm eff} 
= {i\bar\Sigma^> \over 2 \sqrt{s}\Gamma \, {\cal A}\,(1-K)} \,S_{\rm eff}^< 
= {1 \over {\cal A}\,(1-K)} \,S_{\rm eff}^< 
\end{equation}
from which the life time can be immediately read off:
\begin{equation}
  \label{eq:lifetime}
\tau = 2 p_0 \, {\cal A}\,(1-K)  \,.
\end{equation}
As an example the life time of the $\Delta$ baryon is given in Fig.~\ref{fig:tau}
(full line).
The expression on the r.h.s.~can be rewritten by introducing the phase shift
$\delta$ via
\begin{equation}
  \label{eq:defphaseshift}
\tan \delta = {-\sqrt{s}\Gamma \over p^2 - m^2 - {\rm Re}\bar\Sigma^{\rm ret} } \,.
\end{equation}
One finds
\begin{equation}
  \label{eq:taudelta}
\tau = {\partial \delta \over \partial p_0}  \,.
\end{equation}
$\delta$ is displayed in Fig.~\ref{fig:phase}. It shows the typical behavior
of a resonant scattering phase shift (here of pion-nucleon scattering), 
i.e.~it is small for low invariant masses and
rises strongly in the vicinity of the resonance peak mass. There it becomes 
$90^\circ$. 

Relation (\ref{eq:taudelta}) agrees with the one found by Danielewicz and 
Pratt \cite{DanPratt}
but is in striking disagreement with the one commonly used in simulations of 
nucleus-nucleus collisions (see also the discussion in \cite{bass98}), namely
\begin{equation}
  \label{eq:taugamma}
\tau \stackrel{?}{=} {p_0 \over \sqrt{s}\,\Gamma}  \,.
\end{equation}
(Note that in the rest frame of the resonance the latter expression reduces to
$1/\Gamma$ while the r.h.s.~of (\ref{eq:lifetime}) becomes 
$2\sqrt{s} {\cal A}\,(1-K)$.) 

Relation (\ref{eq:taudelta}) can be easily understood if one considers the 
scattering of a wave packet on a potential \cite{GoldWat}: At large distances
$r$ from the interaction region the scattered wave is given by
\begin{equation}
  \label{eq:scawavpac}
\int \!\! d^3 k \, \psi(\vec k - \vec p) \, {e^{ikr} \over r} \, f(E_k,\theta_k) \, 
e^{-i E_k t}
\end{equation}
where $\psi$ denotes the wave packet amplitude. For an amplitude sharply peaked
near $\vec k = \vec p$ one might expand the scattering amplitude $f$ around
$E_p$:\footnote{For simplicity we suppress the angular dependence.}
\begin{eqnarray}
f(E_k) &=& \exp[\ln f(E_k)] \approx 
\exp\left[\ln f(E_p) + (E_k - E_p){\partial \over \partial E_p} \ln f(E_p) \right]
\nonumber \\   \label{eq:expscamp}
&=&  f(E_p) \, 
\exp\left[(E_k - E_p){\partial \over \partial E_p} \ln f(E_p) \right] \,.
\end{eqnarray}
Hence the time delay which the scattered wave packet suffers is given by
\begin{equation}
  \label{eq:timedelay}
{\rm Im}\left[ {\partial \over \partial E_p} \ln f(E_p) \right] = 
{\partial \delta \over \partial E_p}
\end{equation}
where we have introduced the phase shift (for a single channel with angular 
momentum $l$) via
\begin{equation}
  \label{eq:fdel}
f = {1 \over p} \sin\delta \, P_l(\cos\theta_p) \, e^{i\delta}  \,.
\end{equation}
Suppose now that two particles scatter by forming a resonance. In this case
the time delay caused by their scattering can be identified with the life time of 
the formed resonance. This is exactly the essence of relation (\ref{eq:taudelta}).

Both expressions (\ref{eq:lifetime}) and (\ref{eq:taugamma}) are displayed in 
Fig.~\ref{fig:tau}. Obviously their shapes are completely different from each 
other. The most striking difference is their respective threshold behavior.
While $\tau$ as given in (\ref{eq:lifetime}) vanishes at threshold, in contrast
$1/\Gamma$ diverges there as the width vanishes. In addition, $\tau$ is peaked in 
the vicinity of the nominal resonance mass. There $\tau$ is roughly twice as
large as $1/\Gamma$. The latter quantity decreases monotonically in that region.
In spite of their completely different shapes, it is interesting to realize that 
there are intuitive interpretations for 
{\em both} expressions (\ref{eq:lifetime}) and (\ref{eq:taugamma}). 
If a resonance with arbitrary invariant mass is considered
as a ``real particle'', it is clear that its life time increases near
threshold due to the limited phase-space available for the decay products. This
consideration suggests formula (\ref{eq:taugamma}) corresponding to the dashed 
line in Fig.~\ref{fig:tau}. If, however, the resonance is considered as a 
quantum mechanical transient state the uncertainty principle applies. Hence the
more the invariant mass of the resonance deviates from its pole mass the less time
it is allowed to live. Qualitatively this is the essence of (\ref{eq:lifetime}) 
displayed by the full line in Fig.~\ref{fig:tau}. Clearly our derivation given above
supports the latter interpretation, in spite of the fact that what we are 
aiming at is the appropriate life time of (resonance) test-particles which 
intuitively are regarded to be closer to real particles than to quantum mechanical 
states. Nonetheless the transport equation (\ref{eq:transport2}) derived 
from the underlying quantum field theory clearly demands that the appropriate
life time is given by (\ref{eq:lifetime}) instead of (\ref{eq:taugamma}).

At this stage some words of clarification are necessary. For a sharp resonance
it is well-known (e.g.~\cite{PesSchr}) 
that its ``life time'' is given by $1/\Gamma_{\rm on-shell}$
(in its rest frame) which seems to be 
in line with (\ref{eq:taugamma}) but in contrast to 
(\ref{eq:lifetime}) which roughly yields twice as much (cf.~Fig.~\ref{fig:tau}). 
This reasoning, however, is misleading. In the preceding paragraphs we studied the 
life time $\tau$ of a resonance with arbitrary invariant mass. In contrast, the
well-known ``life time'' $\langle \tau \rangle$ of a sharp resonance is a quantity
averaged over the allowed energy (or invariant mass) range. As we shall see now
both expressions (\ref{eq:lifetime}) and (\ref{eq:taugamma}), albeit completely
different in shape, yield the same average value, if the width is sufficiently
small. The natural choice for the probability (density) to find a resonance with
energy $p_0$ is given by the spectral function (\ref{eq:defspec}). 
Recalling the normalization (\ref{eq:norm}) we define
\begin{equation}
  \label{eq:tauav}
\langle \tau \rangle := 
\int\limits_0^{\infty}  {d(p_0^2) \over \pi} \, {\cal A} \, \tau 
\,.
\end{equation}
To evaluate this average for both expressions (\ref{eq:lifetime}) and 
(\ref{eq:taugamma}) we restrict ourselves to the case of small width, i.e.~a sharply
peaked resonance. In this case we can neglect any energy dependence of the width
and simply evaluate it at the pole mass:
\begin{equation}
  \label{eq:gamonsh}
\Gamma_{\rm on-shell} = \Gamma(p_0 = E_p,\vec p)  
\end{equation}
with $E_p := \sqrt{m^2+\vec p^2}$.
For simplicity we also neglect the real part of the retarded self-energy. 
As already pointed out in \cite{leupold00} the following relations hold
(translated here to a relativistic system):
\begin{equation}
  \label{eq:quasipspec}
\left.
{ {\cal A} \atop  2\sqrt{s} \Gamma \, {\cal A}^2} 
\right\}
\to  
\pi \,\delta(p^2 -m^2) \,{\rm sgn}(p_0)
\quad \mbox{for} \quad \Gamma,\, {\rm Re}\bar \Sigma^{\rm ret} \to 0  \,.
\end{equation}
Hence we find for the expression (\ref{eq:lifetime})
\begin{equation}
  \label{eq:tauav2}
\langle \tau \rangle = \int\limits_0^{\infty}  {d(p_0^2) \over \pi} \, 
\underbrace{2\sqrt{s} \Gamma \, {\cal A}^2 \, (1-K)}_{\displaystyle 
\to \pi \,\delta(p^2 -m^2) } 
\, {p_0 \over \sqrt{s} \Gamma}
\to {E_p \over m \Gamma_{\rm on-shell}} \,.
\end{equation}
We get the same result using (\ref{eq:taugamma})
\begin{equation}
  \label{eq:tauav3}
\left\langle {p_0 \over \sqrt{s}\,\Gamma} \right\rangle = 
\int\limits_0^{\infty}  {d(p_0^2) \over \pi} \, 
{\cal A} \, {p_0 \over \sqrt{s}\,\Gamma} \to {E_p \over m \Gamma_{\rm on-shell}} \,.
\end{equation}
In the rest frame of the resonance this reduces to the well-known 
expression\footnote{It is interesting to mention that even for the exemplifying 
case of the $\Delta$ baryon displayed in the figures with its quite sizable on-shell
width of $120\,$MeV the average values in (\protect\ref{eq:tauav2}) and 
(\protect\ref{eq:tauav3}) do not deviate much from each other: 
$\langle \tau \rangle \approx 1.68\,$fm/c, 
$\langle 1/\Gamma \rangle \approx 1.55 \,$fm/c, 
$1/\Gamma_{\rm on-shell} \approx 1.64\,$fm/c.}
\begin{equation}
  \label{eq:tauav4}
\langle \tau \rangle = 1/\Gamma_{\rm on-shell}  \,.
\end{equation}
Note, however, that this relation provides only an average value of the life time. 
It does not answer the question how long a resonance with arbitrary 
invariant mass lives. 

Concerning practical applications, i.e.~transport simulations
with test-particles, the previous considerations suggest the following 
two possibilities for states with a small width: 
\begin{itemize}  \label{page:onoff}
\item[1.] One might represent $S_{\rm eff}^<$
by test-particles having arbitrary invariant masses. For small width the likelihood
to produce a test-particle with a mass far away from the pole mass is of course
negligibly small. Therefore basically all test-particles will have similar (but not
exactly the same) invariant masses. The life time of a test-particle is given by
(\ref{eq:lifetime}). 
\item[2.] The second possibility is to 
attribute the on-shell mass to all test-particles. In this case the test-particles 
represent the integrated quantity $\int d(p_0^2)\,S_{\rm eff}^<$ instead of
$S_{\rm eff}^<$. Consequently all test-particles have to decay with the same 
integrated
(averaged) life time $\langle \tau \rangle = 1/\Gamma_{\rm on-shell}$. 
\end{itemize}
Obviously, concerning states with a large width only the first possibility is
reasonable. From a practical point of view a word of caution is necessary. To 
realize the first possibility one has to make sure that one has chosen enough
test-particles to ``fill out'' the whole spectral distribution. Especially, if the 
test-particles in the tails of the spectral distribution are missing one might 
obtain an average life time which is too high since (cf.~Fig.~\ref{fig:tau})
\begin{equation}
  \label{eq:peakandav}
\tau_{s \approx m^2} \approx 2/\Gamma_{\rm on-shell} \approx
2 \, \langle \tau \rangle  \,.
\end{equation}
Hence for states with a small width the second possibility described above
seems to be more practical. 

\begin{figure}[htbp]
\centerline{\psfig{figure=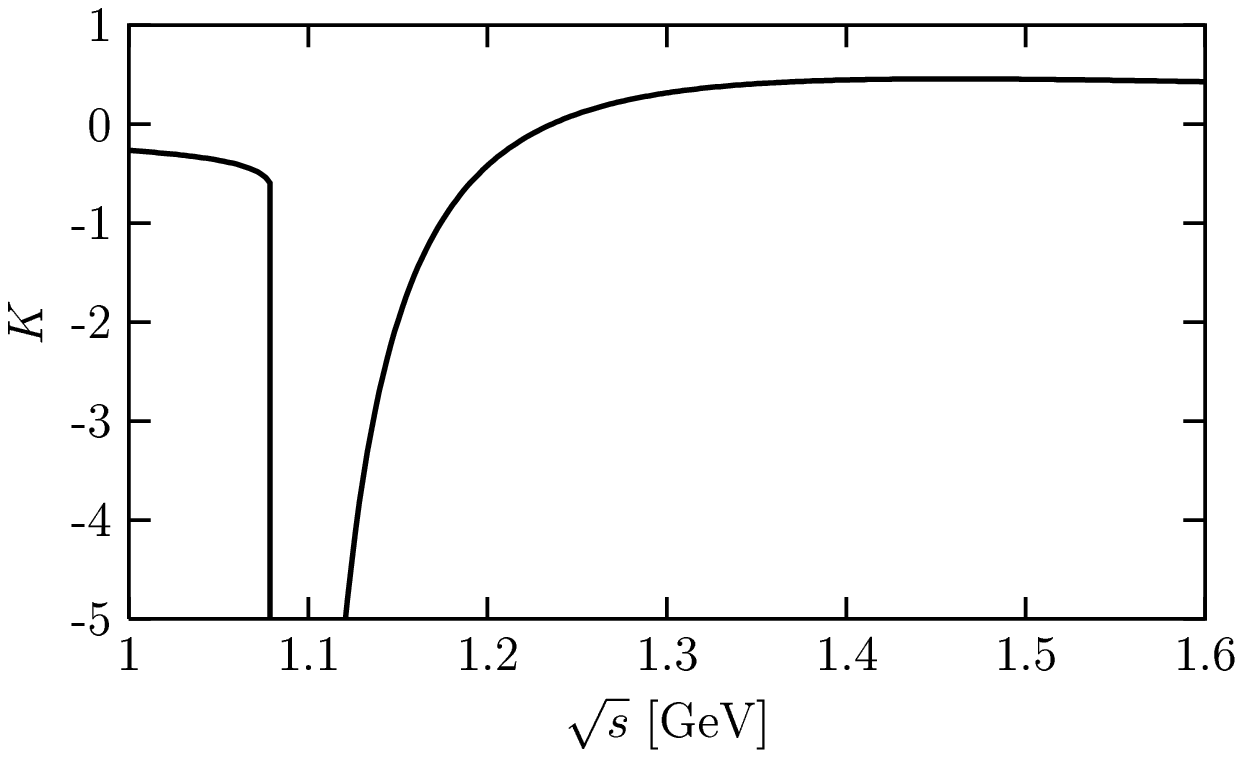}}
\caption{$K$ as given in (\protect\ref{eq:defk})
as a function of the invariant mass. }
\label{fig:kfac}
\end{figure}

\begin{figure}[htbp]
\centerline{\psfig{figure=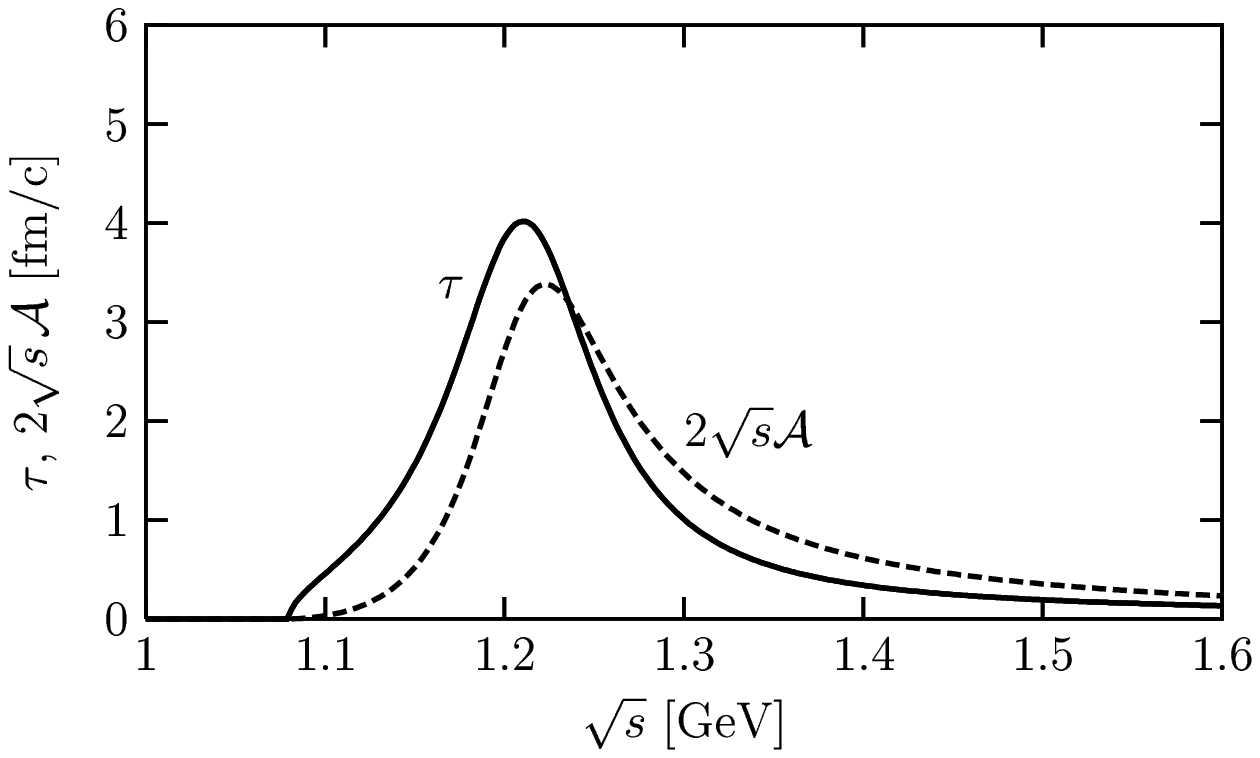}}
\caption{Life time $\tau$ (full line) as given in (\protect\ref{eq:lifetime})
and $2 \sqrt{s} \, {\cal A}$ (dashed line)
as functions of the invariant mass. Note that $\tau$ is evaluated in the rest 
frame of the decaying resonance, i.e.~$p_0 = \sqrt{s}$.}
\label{fig:tausimple}
\end{figure}

It is worth to discuss the role of the factor $(1-K)$ in the expression for the
life time (\ref{eq:lifetime}). Due to the contribution (cf.~(\ref{eq:defk}))
\begin{equation}
  \label{eq:difterm}
{1 \over \sqrt{s}\Gamma}
{\partial(\sqrt{s}\Gamma) \over \partial p_0}
\end{equation}
$K$ diverges at threshold, i.e.~where $\Gamma$ vanishes. This is shown in 
Fig.~\ref{fig:kfac}. We also deduce from that figure that $(1-K)$ remains positive
for all values of $s$. This is a necessary consistency condition since the life time
and also the effective particle number density (\ref{eq:defskleff}) should not 
become negative (cf.~the corresponding discussion in \cite{leupold00}). 
In Fig.~\ref{fig:tausimple} we compare the life time $\tau$ from (\ref{eq:lifetime})
with the simpler quantity $2 \sqrt{s} \, {\cal A}$. As can be seen the main effect
of $(1-K)$ on $\tau$ is to change the threshold behavior of the spectral function
for the $\Delta$ (cf.~(\ref{eq:gamdelta},\ref{eq:defbeta}))
\begin{equation}
  \label{eq:specthres}
{\cal A} \sim \Gamma \sim k_{\rm rel}^3 \sim 
\left(\sqrt{s}-(m_N+m_\pi) \right)^{3/2}
\qquad \mbox{for} \quad \sqrt{s} \approx m_N+m_\pi
\end{equation}
to
\begin{equation}
  \label{eq:tauthres}
\tau \sim {\cal A}\,(1-K) \sim \left(\sqrt{s}-(m_N+m_\pi) \right)^{1/2}
\qquad \mbox{for} \quad \sqrt{s} \approx m_N+m_\pi  \,.
\end{equation}
This explains the kink of $\tau$ at threshold as compared to the smooth rising
of the spectral function. 
In total, however, $\tau$ and $2 \sqrt{s} \, {\cal A}$ look rather similar,
especially if compared to the completely different shape of $1/\Gamma$ displayed in
Fig.~\ref{fig:tau}.

\subsection{More than one decay channel}

We now turn to an extension of the formalism applicable when there exists more than 
one decay channel for the resonance. While the {\em total} life time is given by 
(\ref{eq:lifetime}) it is also important to know the {\em partial} decay rates 
(inverse life times) for the various channels. Formally this can be accounted for
by realizing that $\bar\Sigma^> =: \bar\Sigma^>_{\rm tot}$ is given by a sum of 
the self-energies corresponding to the different decay channels:
\begin{equation}
  \label{eq:splitdecayrate}
i\bar\Sigma^>_{\rm tot} = \sum\limits_j i\bar\Sigma^>_j  
= \sum\limits_j 2\sqrt{s}\Gamma_j  \,.
\end{equation}
The latter equality makes sense if there are no production channels, 
i.e.~$\bar\Sigma^<_j = 0$. Now it is straightforward to deduce from 
(\ref{eq:timeevsimple}) the evolution equation
\begin{equation}
  \label{eq:tsimpmore}
\partial_t S^<_{\rm eff} = - \sum\limits_j {1 \over \tau_j} S^<_{\rm eff}
\end{equation}
with the partial life times 
\begin{equation}
  \label{eq:partlifetime}
\tau_j = {2 p_0 \, \Gamma_{\rm tot} \, {\cal A}\,(1-K) \over \Gamma_j}  
= {\Gamma_{\rm tot} \over \Gamma_j} \tau_{\rm tot}  
\end{equation}
where we have attributed the index ``tot'' to all expressions calculated earlier
to distinguish them from the expressions belonging to a single one of the decay 
channels. We note in passing that the probability for the resonance to decay
into channel $j$ (branching ratio) agrees with the probability that one would 
deduce from (\ref{eq:taugamma}):
\begin{equation}
  \label{eq:probdec}
{\tau_{\rm tot} \over \tau_j} = {\Gamma_j \over \Gamma_{\rm tot} } =
{{ \displaystyle 1 \over \displaystyle \Gamma_{\rm tot}} \over 
{\displaystyle 1 \over \displaystyle \Gamma_j } }  \,.
\end{equation}

\subsection{In-medium case}

So far, we have restricted our considerations to resonance decays in vacuum
neglecting the production channels encoded in $\bar\Sigma^<$. If we want to
study creation and decay of resonances in a medium, we have to take both loss 
and gain terms, i.e.~the full r.h.s.~of (\ref{eq:transport2}) into account.
For simplicity we ignore the in general rather lengthy l.h.s.~of 
(\ref{eq:transport2}) which describes the propagation of the test-particles
according to drift term, Vlasov term, etc.
For our purposes it is enough to study the simplified evolution equation
\begin{equation}
  \label{eq:transport3}
-2p_0 \, \partial_t S_{\rm eff}^< = 
{i\bar\Sigma^> \over 2 \sqrt{s}\Gamma \, {\cal A}\,(1-K)} \,S_{\rm eff}^< 
- i\bar\Sigma^<\,S^>  \,.
\end{equation}
In general, $i\bar\Sigma^>$ and $i\bar\Sigma^<$ are connected by detailed balance.
For the case of thermal equilibrium this leads to the Kubo-Martin-Schwinger (KMS)
boundary condition (cf.~e.g.~\cite{GrLe98})
\begin{equation}
  i\bar \Sigma^<(p) = i\bar \Sigma^>(p) \, e^{-p_0/T}  \,.   
\label{eq:heat}
\end{equation}
To simplify the discussion and to get a closer relation to the life time 
introduced above we assume in the following 
\begin{equation}
  \label{eq:lmsg}
i\bar \Sigma^<(X,p) \ll i\bar \Sigma^>(X,p) 
\end{equation}
(concerning the thermal state we turn from a Bose
to a Boltzmann distribution). Using (\ref{eq:defwidth}) we find 
\begin{equation}
  \label{eq:boltz1}
i\bar \Sigma^>(X,p) \approx 2\sqrt{s}\Gamma(X,p)   \,.
\end{equation}
The KMS-condition (detailed balance) becomes
\begin{equation}
  \label{eq:boltz2}
i\bar \Sigma^<(p) \approx 2\sqrt{s}\Gamma(p) \, e^{-p_0/T}  
\end{equation}
while (\ref{eq:transport3}) simplifies to
\begin{equation}
  \label{eq:collsimple}
\partial_t S_{\rm eff}^< =
-{1 \over \tau} \,S_{\rm eff}^< + {i\bar\Sigma^< \over p_0} \, {\cal A}
\end{equation}
where we have replaced $b$ as defined in (\ref{eq:bose}) by 1.
Hence we deduce that the effective loss rate is given by $1/\tau$
while the gain rate is $i\bar\Sigma^</p_0$. However, since $\sqrt{s}\Gamma(p)/p_0$ 
and not $1/\tau$ is connected to $i\bar\Sigma^</p_0$ by detailed balance 
(\ref{eq:boltz2}) one
might worry about the correct thermal limit for the particle number density.
Thermal equilibrium, i.e.~a stationary solution for $S_{\rm eff}^<$ is reached if 
gain and loss terms cancel each other:
\begin{equation}
  \label{eq:gaineqloss}
-{1 \over \tau} \,S_{\rm eff}^< + {i\bar\Sigma^< \over p_0} \, {\cal A}
\to 0  \,.
\end{equation}
Using the definition of the life time (\ref{eq:lifetime}) and the (simplified)
KMS-condition (\ref{eq:boltz2}) we end up with
\begin{equation}
  \label{eq:boltz3}
S_{\rm eff}^< \to 4\sqrt{s}\Gamma \, {\cal A}^2 \, (1-K) \, e^{-p_0/T}
\approx 4\sqrt{s}\Gamma \, {\cal A}^2 \, (1-K) \, n_B
\end{equation}
where we have replaced the Boltzmann by the Bose function in the last 
step.\footnote{Note that an exact treatment of the r.h.s.~of 
(\protect\ref{eq:transport3}) together with the exact KMS-condition 
(\protect\ref{eq:heat}) would have led immediately to the last expression of 
(\protect\ref{eq:boltz3}). We made the approximations only to get closer contact
to the expression for the life time.}
Indeed this is the correct thermal limit for $S_{\rm eff}^<$ as already presented
in (\ref{eq:thermal}). Hence we have shown that the combination of loss and gain
terms on the r.h.s.~of the transport equation (\ref{eq:transport2}) indeed 
yields the correct thermal limit for the effective particle number density 
$S^<_{\rm eff}$. 
Turning the argument around we conclude that attributing the life time 
(\ref{eq:lifetime}) to a
resonance only leads to thermodynamically consistent results if one realizes that 
the appropriate
(effective) particle number density is given by $S^<_{\rm eff}$ introduced in 
(\ref{eq:defskleff}) and {\em not} by $S^<$. Assuming erroneously that the
test-particles which decay according to (\ref{eq:lifetime}) represent $S^<$
one would essentially solve an evolution equation like 
\begin{equation}
  \label{eq:wrong}
\partial_t S^< = - {1 \over \tau} \,S^< + {i\bar\Sigma^< \over p_0} \, {\cal A}
\qquad \mbox{(wrong!)}
\end{equation}
instead of (\ref{eq:collsimple}). The thermal limit of the (wrong) equation 
(\ref{eq:wrong}) is of course given by
\begin{equation}
  \label{eq:wrong2}
S^< \to 4\sqrt{s}\Gamma \, {\cal A}^2 \, (1-K) \, e^{-p_0/T}
\qquad \mbox{(wrong!)}
\end{equation}
in contrast to the correct behavior (cf.~(\ref{eq:thermal}))
\begin{equation}
  \label{eq:right}
S^< \to 2 {\cal A}\, n_B  \approx  2 {\cal A}\, e^{-p_0/T}   \,.
\end{equation}
To conclude we have deduced the expression (\ref{eq:lifetime}) for the life time 
from the transport equation (\ref{eq:transport2}) for $S^<_{\rm eff}$. 
In turn we have also shown
that starting from the life time (\ref{eq:lifetime}) for test-particles one has
to realize that these test-particles represent $S^<_{\rm eff}$ 
and not $S^<$.\footnote{Concerning these detailed balance considerations the use of 
$1/\Gamma$ for the life time would be compatible with $S^<$.}


\section{Summary}   \label{sec:sum}

After a short review of the derivation of a transport equation from the underlying
quantum field theoretical Kadanoff-Baym equations we have focused on the 
test-particle representation of the transport process.  
At present this seems to be the
most promising tool to solve the transport equation in practice. We have pointed
out (repeating the arguments of \cite{leupold00}) that the proper quantity which
should be simulated by test-particles, i.e.~which counts the test-particles per
four-momentum and space volume, is given by the effective particle number density
$S^<_{\rm eff}$ introduced in (\ref{eq:defskleff}). Once this
identification has been made it was straightforward to derive from the 
transport equation (\ref{eq:transport2}) the life time of the resonance 
test-particles. For the vacuum case we obtained expression (\ref{eq:lifetime})
which can be rewritten as (\ref{eq:taudelta}) confirming
the result of Danielewicz and Pratt \cite{DanPratt}. As can be 
seen from Fig.~\ref{fig:tau} this result for the life time strongly differs from 
the inverse width prescription commonly used in transport simulations. If more than
one decay channel for the resonance has to be considered, the partial life time
is given by (\ref{eq:partlifetime}). The latter formula also applies if (in a 
medium) collisional broadening modifies the total width. Note that for
the calculation of the life times the required spectral function as well as total 
and partial widths should be determined self-consistently for the in-medium case.
Finally we have shown that
the identification of the proper effective particle number density 
(\ref{eq:defskleff}) and the relation for the life time (\ref{eq:lifetime})
are intimately connected. On the one hand, the life time formula 
(\ref{eq:lifetime}) is derived from the transport equation (\ref{eq:transport2}) 
for the effective particle number density (\ref{eq:defskleff}). On the other 
hand, starting out from the formula for the life time and assuming thermal
equilibrium leads to an expression for the particle number density which agrees
with the thermodynamic limit of $S^<_{\rm eff}$.

In \cite{leupold00} it has been shown that the scattering cross sections for 
off-shell test-particles\footnote{Here ``off-shell test-particles'' denote 
test-particles with arbitrary invariant mass as they appear in the definition of
$f$ in (\protect\ref{eq:testpart}). By chance such test-particles even 
might be on their mass-shell --- still we would call them ``off-shell 
test-particles''. This phrase merely denotes a concept than a single particle, 
namely the concept that ``off-shell test-particles'' fill the generalized 
phase-space of {\em four}-momentum and coordinate space (weighted by their spectral 
distribution). In contrast ``on-shell test-particles''
live in usual phase-space. While ``off-shell test-particles'' represent
$f$ ``on-shell test-particles'' represent
$\int d(p_0^2) f$. These concepts are also discussed at page \pageref{page:onoff}
(items 1.~and 2.).} have to be modified to account for the fact that these 
test-particles represent the effective particle number density $S^<_{\rm eff}$ and 
not $S^<$. There are rather simple rules for the required modification: 
\begin{itemize}
\item[1.] For each incoming off-shell test-particle the cross section has to be
divided by 
\begin{equation}
  \label{eq:renormf}
r = 2 \sqrt{s}\Gamma \, {\cal A}\,(1-K)
\end{equation}
where all quantities refer to the (in-medium) properties of the off-shell 
test-particle, i.e.~$\Gamma$ is its total (in-medium) width etc.
\item[2.] For each outgoing off-shell test-particle the Pauli blocking/Bose
enhancement factor is given by (cf.~(\ref{eq:bose}))
\begin{equation}
  \label{eq:pauliblock}
1 \mp {S_{\rm eff}^< \over 4 \sqrt{s}\Gamma \, {\cal A}^2\,(1-K)}   \,.
\end{equation}
\end{itemize}
In the same way the result for the life time can be understood: In complete analogy
to the cross section the decay width $\Gamma_j$ has to be divided by 
(\ref{eq:renormf}). In that way the inverse of that quantity changes from
$1/\Gamma_j$ to (\ref{eq:partlifetime}) which is our expression for the
life time. 
Whether these new rules for the calculation of cross sections and decay 
probabilities lead to observable effects in nucleus-nucleus collisions remains
to be seen.

\acknowledgements I acknowledge fruitful discussions with J.~Aichelin, V.~Koch,
and U.~Mosel during the ``Hirschegg 2000'' workshop. These discussions initiated
the work presented here. I also thank W.~Cassing, C.~Greiner, and S.~Juchem for
their helpful comments on the manuscript.

\begin{appendix}

\section{A simple model for the $\Delta$ resonance}

To exemplify the formulae given above we study a toy model for the $\Delta$
resonance where spin (and also isospin) is neglected for simplicity. The width is 
determined by the $\Delta \to N\,\pi$ decay channel:
\begin{equation}
  \label{eq:gamdelta}
\Gamma(s) = \Gamma_0 {\beta(s) \over \beta(m_\Delta^2)}  \qquad 
\mbox{for $\sqrt{s} \ge m_N + m_\pi$}
\end{equation}
with
\begin{equation}
  \label{eq:defbeta}
\beta(s) = {k_{\rm rel} \over s} {(k_{\rm rel} R)^2 \over 1 + (k_{\rm rel} R)^2 } 
\,.
\end{equation}
Here the momentum $k_{\rm rel}$ of $\pi$ and $N$ in the rest frame of the decaying 
$\Delta$ resonance with invariant mass $\sqrt{s}$ is given by
\begin{equation}
  \label{eq:defkrel}
k_{\rm rel} = { \left[
\left(s-(m_N+m_\pi)^2\right) \, \left(s-(m_N-m_\pi)^2\right) \right]^{1/2}
\over 2 \sqrt{s} }  \,.
\end{equation}
In (\ref{eq:defbeta}) we have included a Blatt-Weisskopf function \cite{Blattweiss}
parametrized
by the length parameter $R$. We adopt the following values to simulate the vacuum
decay properties of the $\Delta$ resonance:
$m_N = 940\,$MeV, $m_\pi = 140 \,$MeV, $m_\Delta = 1232\,$MeV, 
$\Gamma_0 = 120\,$MeV, and $R = 1 \,$fm.

Since $\sqrt{s}\Gamma$ becomes constant for large energies the real part of the
retarded self-energy can be calculated by a one-time subtracted dispersion relation
according to (\ref{eq:dispersion}). We choose the subtraction constant such that
the real part of the self-energy vanishes for $s = m_\Delta^2$. In our simple
vacuum model the energy integration in (\ref{eq:dispersion}) can be rewritten
in terms of the invariant mass squared. One finally gets:
\begin{equation}
  \label{eq:dispexa}
{\rm Re}\bar\Sigma^{\rm ret}(s) = -(s-m_\Delta^2) \;\;
{\cal P} \!\! \int\limits_{(m_N + m_\pi)^2}^\infty \! {ds' \over \pi} 
{ \sqrt{s'} \Gamma(s')  \over (s-s')(m_\Delta^2-s') }   \,.
\end{equation}

\end{appendix}


\end{document}